\documentclass{aastex}
\usepackage{spr-astr-addons}
\usepackage{url}
\usepackage{scrextend}
\usepackage{float}
\usepackage[multiple]{footmisc}

\usepackage{threeparttable}
\usepackage{graphicx}
\RequirePackage{color}

\newcommand{\n}{\mbox {N\,103B}}
\def\HI{\hbox{H\,{\sc i}}}
\def\HII{\hbox{H\,{\sc ii}}}
\def\p0{\phantom{0}}

\def\arcmin{\hbox{$^\prime$}}
\def\arcsec{\hbox{$^{\prime\prime}$}}
\def\degr{\hbox{$^\circ$}}

\def\farcs{\hbox{$.\!\!^{\prime\prime}$}}

\newcommand{\kms}{km s$^{-1}$}

\def\msun{$M_{\odot}$}

\begin{document}
\title{Radio Emission from Interstellar Shocks: \\
Young Type\,Ia Supernova Remnants and the Case of \n\ in the Large Magellanic Cloud \\ 
}
\slugcomment{Not to appear in Nonlearned J., 45.}
\shorttitle{Radio Emission from Interstellar Shocks: A Young Type\,Ia LMC SNR \n \\}
\shortauthors{Alsaberi et al.}

\author{Alsaberi R. Z. E.\altaffilmark{1}} 
\and \author{Barnes L. A.\altaffilmark{1}} 
\and \author{Filipovi\'c M. D.\altaffilmark{1}} 
\and \author{Maxted N. I.\altaffilmark{1,2}} 
\and \author{Sano H.\altaffilmark{3}}
\and \author{Rowell G.\altaffilmark{4}} 
\and \author{Bozzetto L. M.\altaffilmark{1}} 
\and \author{Gurovich S.\altaffilmark{5}}
\and \author{Uro\v sevi\' c D.\altaffilmark{6}} 
\and \author{Oni\' c D.\altaffilmark{6}}
\and \author{For B.-Q.\altaffilmark{7}} 
\and \author{Manojlovi\'c P.\altaffilmark{1,8}} 
\and \author{Wong G.\altaffilmark{1}} 
\and \author{Galvin T.\altaffilmark{1,8}}
\and \author{Kavanagh P.\altaffilmark{9}}
\and \author{Ralph N.\altaffilmark{1}}
\and \author{Crawford E. J.\altaffilmark{1}}
\and \author{Sasaki M.\altaffilmark{10}}
\and \author{Haberl F.\altaffilmark{11}}
\and \author{Maggi P.\altaffilmark{12}}
\and \author{Tothil N. F. H.\altaffilmark{1}}
\and \author{Fukui Y.\altaffilmark{3}}

\affil{$^1$Western Sydney University, Locked Bag 1797, Penrith South DC, NSW 1797, Australia}
\affil{$^2$School of Science, University of New South Wales, Australian Defence Force Academy, Canberra, ACT 2600, Australia}
\affil{$^3$Institute for Advanced Research, Nagoya University, Chikusa-ku, Nagoya 464-8601, Japan}
\affil{$^4$School of Physical Sciences, University of Adelaide, North Terrace, Adelaide, SA 5005, Australia}
\affil{$^5$Instituto de Astronom\'ia Te\'orica y Experimental - Observatorio Astron\'omico Co\'rdoba (IATE-OAC-UNC-CONICET)}
\affil{$^6$Department of Astronomy, Faculty of Mathematics, University of Belgrade, Studentski trg 16, 11000 Belgrade, Serbia}
\affil{$^7$International Centre for Radio Astronomy Research, University of Western Australia, 35 Stirling Hwy, Crawley, WA 6009, Australia. ARC Centre of Excellence for All Sky Astrophysics in 3 Dimensions (ASTRO 3D)
\affil{$^8$CSIRO Astronomy and Space Sciences, Australia Telescope National Facility, PO Box 76, Epping, NSW 1710, Australia}
\affil{$^{9}$School of Cosmic Physics, Dublin Institute for Advanced Studies, 31 Fitzwillam Place, Dublin 2, Ireland}
\affil{$^{10}$Remeis Observatory and ECAP, Universit\"{a}t Erlangen-N\"{u}rnberg, Sternwartstr. 7, 96049 Bamberg, Germany}
\affil{$^{11}$Max-Planck-Institut f\"{u}r extraterrestrische Physik, Giessenbachstra\ss e, 85748 Garching, Germany}
\affil{$^{12}$Universit\'e de Strasbourg, CNRS, Observatoire astronomique de Strasbourg, UMR 7550, F-67000 Strasbourg, France}
}



\begin{abstract}

We investigate young type\,Ia supernova remnants (SNRs) in our Galaxy and neighbouring galaxies in order to understand their properties and early stage of their evolution. Here we present a radio continuum study based on new and archival data from the Australia Telescope Compact Array (ATCA) towards \n, a young ($\leq$1000\,yrs) spectroscopically confirmed type\,Ia SNR in the Large Magellanic Cloud (LMC) and proposed to have originated from a single degenerate (SD) progenitor. The radio morphology of this SNR is asymmetrical with two bright regions towards the north-west and south-west of the central location as defined by radio emission. 

\n\ identified features include: a radio spectral index of --0.75$\pm$0.01 (consistent with other young type\,Ia SNRs in the Galaxy); a bulk SNR expansion rate as in X-rays; morphology and polarized electrical field vector measurements where we note radial polarisation peak towards the north-west of the remnant at both 5500 and 9000\,MHz. The spectrum is concave-up and the most likely reason is the non-linear diffusive shock acceleration (NLDSA) effects or presence of two different population\textbf{s} of ultra-relativistic electrons.

We also note unpolarized clumps near the south-west region which is in agreement with this above scenario. We derive a typical magnetic field strength for \n\, of 16.4\,$\mu$G for an average rotation measurement of 200\,rad\,m$^{-2}$. However, we estimate the equipartition field to be of the order of $\sim$235\,$\mu$G with an estimated minimum energy of E$_{\rm min}$=6.3$\times10^{48}$\,erg. The close ($\sim 0.5$\degr) proximity of \n\ to the LMC mid-plane indicates that an early encounter with dense interstellar medium may have set an important constrain on SNR evolution. 

Finally, we compare features of \n\, to six other young type\,Ia SNRs in the LMC and Galaxy, with a range of proposed degeneracy scenarios to highlight potential differences due to a different models. We suggest that the single degenerate scenario might point to morphologically asymmetric type\,Ia supernova explosions. 



\end{abstract}



\section{Introduction} 
 \label{sec:intro}

Supernovae (SNe) play important roles in cosmology. Type\,Ia SNe have been used as standard candles to constrain cosmological parameters, providing the first and best evidence of cosmic acceleration and dark energy \citep{1998AJ....116.1009R,1999ApJ...517..565P}. Further, SNe shape the way that galaxies form and evolve. SNe distribute the products of stellar nucleosynthesis into the interstellar medium (ISM), enriching, heating and compressing gas that forms second and third generation stars. The combined effect of many SN explosions in star-forming galaxies can drive gas out of galaxies and into the circum- and intergalactic medium. Observations of high-redshift star-forming galaxies have shown abundant evidence of galactic-scale outflows; as \citet{2010ApJ...717..289S} and \citet{2019MNRAS.482.3089N} have noted, virtually every $z>2$ galaxy bright enough to be observed spectroscopically is driving out material at several hundred kilometers per second.

As a result, observations of SNe in the local/nearby Universe can inform both the measurement of cosmological parameters and theoretical models of galaxy formation. Data from nearby SNe and supernova remnants (SNRs) via systematic monitoring campaigns are particularly useful since many of these systems offer time-domain data that are costly to obtain or not possible at higher redshift.

In particular, nearby Universe observations have the potential for understanding  the causes of SNe, assuming that the explosion mechanisms do not significantly change over cosmic time. Type\,Ia SNe (a.k.a. thermonuclear (TN)) are believed to occur when a white dwarf (WD) close to the Chandrasekhar mass, reigniting explosive nuclear reactions. This is most likely to occur in a binary system, either by the accretion from a non-degenerate companion star (the so-called single degenerate scenario, SD) or via a merger with another WD (double degenerate scenario, DD). Core collapse SNe (CC; SN\,II, SN\,Ib and SN\,Ic) result from the collapse of a single short lived massive star ($>$~8\msun) as it exhausts its nuclear fuel. Intriguingly, the ratio of SN\,Ia to CC explosions is lower in large disk galaxies (like the Milky Way) than in dwarf irregular galaxies, implying that Galactic environment and different star formation rate (SFR) cannot be ignored \citep{1995MNRAS.277..945T}. This effect is been shown by the recent observational evidence from nearby dwarfs--Magellanic Clouds \citep[see ][]{2016A&A...585A.162M}.

A major remaining unknown is the proportion of type\,Ia SNe that result from the SD and DD scenarios \citep[see e.g.][]{Nomoto:1991,Fryer:2010}. The decisive signature of an SD explosion is the observation of the surviving companion star, because the merger of the two WDs in the DD scenario leaves no remnant star. Other signatures include a dense circumstellar medium \citep{2001A&A...373..281D,2007ApJ...662..998B,2012ApJ...755....3W}, accretion winds from the progenitor system \citep{1996ApJ...470L..97H,2018ApJ...867....7S}, and strong K-shell emission from Ni and Mn \citep{2015ApJ...801L..31Y}. Detailed observations of type\,Ia SNRs, to reconstruct the scene of the explosion and pinpoint the likely location of a surviving companion (if it exists), are required to distinguish between the SN scenarios. There may be other ways in which SNR indicate the scenario of their creation \citep{2016ApJ...819...37C}. 

Studies of SNRs in the Milky Way suffer from imprecise distance measurements and foreground absorption from the Galactic plane. Many of these problems are largely overcome by studying SNRs in the Large Magellanic Cloud (LMC). The LMC, with an distance of $\sim$50\,kpc \citep{mt,2019Natur.567..200P} and a near face on orientation (inclination angle of 35\degr), is near enough to allow deep, high-resolution, multi-frequency observations \citep{2016A&A...585A.162M,2017ApJS..230....2B}. The LMC has active star-forming regions, and is located outside of the Galactic plane where absorption by gas and dust is reasonably low. The Small Magellanic Cloud (SMC), by contrast, has a smaller SNR population \citep{2008A&A...485...63F,2019arXiv190811234M,2019MNRAS.487.4332T,2019MNRAS.tmp.2277J}, and in particular, no obvious young ($<$~1000\,yrs) type\,Ia SNRs \citep{2015ApJ...803..106R,2019ApJ...881...85S}. The nearby Andromeda group (including M\,31 and M\,33) is too distant to allow the study of SN degeneracy scenarios \citep{2012SerAJ.184...41G,2014SerAJ.189...15G,2018A&A...620A..28S}. As a result, LMC SNRs have been the subject of our considerable observational effort \citep{2007MNRAS.378.1237B,2009SerAJ.179...55C,2008SerAJ.177...61C,2010A&A...518A..35C,2014AJ....148...99C,2010SerAJ.181...43B,2012A&A...539A..15G,2012MNRAS.420.2588B,2012RMxAA..48...41B,2012SerAJ.184...69B,2012SerAJ.185...25B,2012A&A...540A..25D,2012A&A...543A.154H,2013A&A...549A..99K,2013MNRAS.432.2177B,2014AJ....147..162D,2014ApJ...780...50B,2014MNRAS.439.1110B,2014MNRAS.440.3220B,2014Ap&SS.351..207B,2014A&A...567A.136W,2014A&A...561A..76M,2015MNRAS.454..991R,2015A&A...573A..73K,2015A&A...579A..63K,2015A&A...583A.121K,2015PKAS...30..149B,2016A&A...586A...4K,2019A&A...621A.138K,2019MNRAS.486.2507A,2019arXiv191002792M}.

\begin{figure*}[ht!]
\centering
\includegraphics[width=\textwidth]{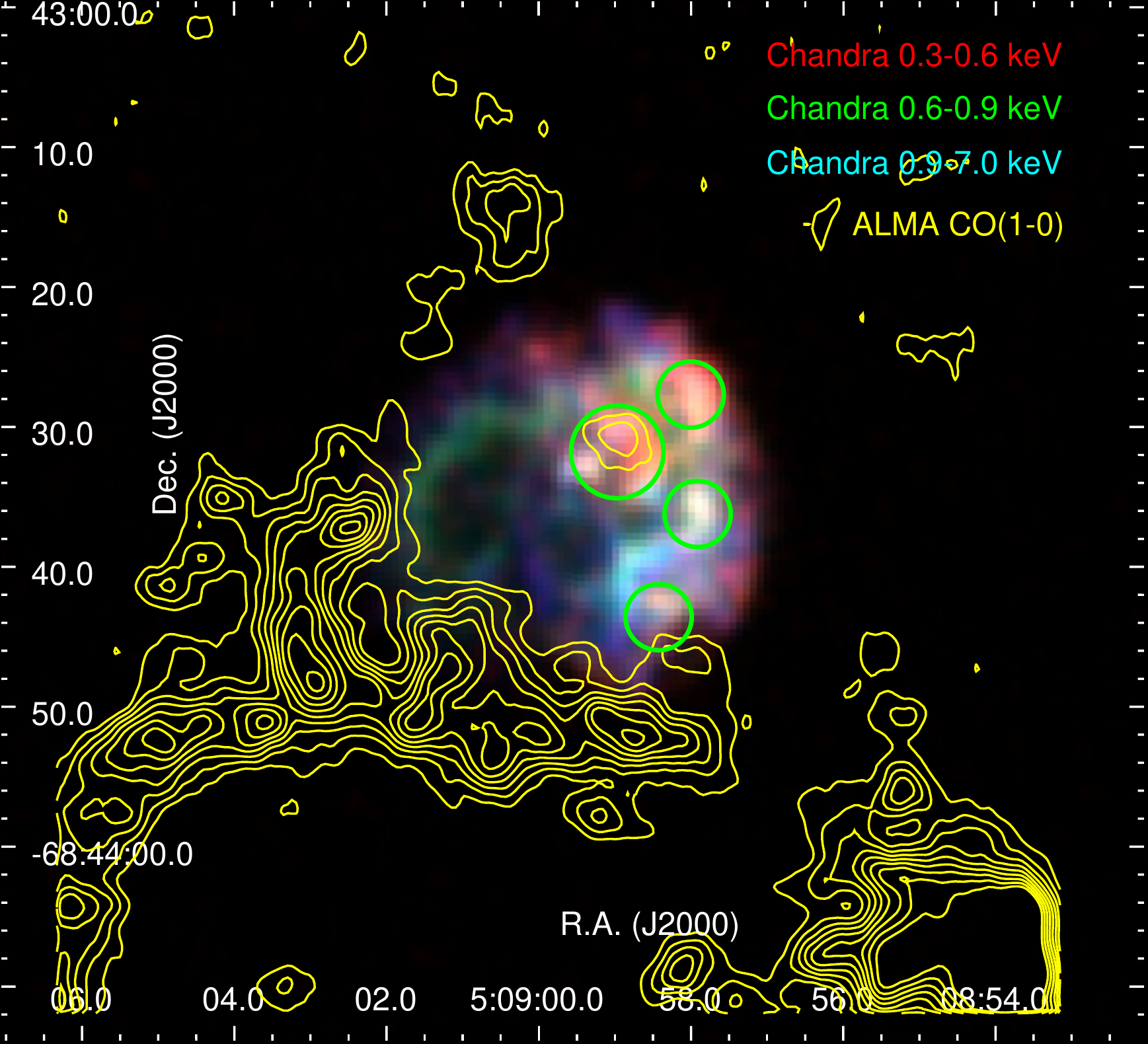}
\caption{Three-colour image of Chandra X-ray emission from \n\ \citep{Lewis:2003}. The colours represent three energy bands: 0.3--0.6\,keV, 0.6--0.9\,keV, and 0.9--7.0\,keV (red, green, and blue, respectively). Yellow contours correspond to ALMA CO(1--0) emission integrated between 244.8 and 252.8\,km\,s$^{-1}$ (contour levels are 5, 10, ..., 45\,K\,km\,s$^{-1}$), and indicate the potential molecular gas association found by \citet{2018ApJ...867....7S}. Four circles show regions containing optical nebula knots, as seen in the HST H$\alpha$ emission \citep{li}. 
\label{XrayFig1}}
\end{figure*}

In this paper, we focus on \n, a young (380--860\,yrs) SNR in the LMC \citep{res}. It is located close to a dense \HII\ region \citep{ck}, and only $\sim40$\,pc from the young star cluster NGC~1850 \citep{li}. X-ray \citep{hu}, CO \citep{2018ApJ...867....7S} and optical studies \citep{res,res2} have shown that it is likely to be a type\,Ia SNR; \cite{li} observed a main sequence star that is likely to be the surviving companion. As shown in Figure~\ref{XrayFig1}, \n\ is not a simple spherical shell, showing significant structural asymmetries. The remnant has a diameter of 6.8\,pc \citep{2017ApJS..230....2B}, with the Western hemisphere that is $\sim3$~times brighter than the Eastern hemisphere \citep{dm}. The shell is Balmer-dominated in the optical wavelengths, which indicates the effects of propagating to a partially neutral circumstellar medium \citep[CSM][]{gh}. \n\ is somewhat similar to the Kepler SNR, another young type\,Ia SNR \citep{2007ApJ...668L.135R}. Recent morphological studies of \HI\ and CO reinforce this picture, and suggest that the SN explosion occurred inside a medium heavily influenced by nearby OB stars and by the SN progenitor itself before it enters the WD stage \citep{2018ApJ...867....7S}.

Our paper is organised as following: Section~\ref{sec:data} introduces our observations and data reduction. Section~\ref{sec:results} shows our polarisation images of \n, discussing its implications for the supernova and its aftermath. In Section~\ref{sec:comparison} we compare \n\ to similar young type\,Ia SNRs and finally conclusions are summarized in Section~\ref{sec:con}.

\section{New ATCA Observations and augmented Data}  
 \label{sec:data}
 
The new and archival \footnote{Australia Telescope Online Archive (ATOA), hosted by the Australia Telescope National Facility (ATNF): atoa.atnf.csiro.au} ATCA observations of \n\, are now discussed. The data from two ATCA observation campaigns: one prior to the Compact Array Broadband Backend (CABB) upgrade (project C148), and the other using CABB (projects CX310, CX403, and C3229). Details of these observations are listed in Table~\ref{tab1}.

\begin{table*}
\caption{ATCA observational data of SNR \n.
\label{tab1}}
\begin{tabular}{@{}crrrcrl@{}}
\tableline
Date         &Project& Array         & No.      & Bandwidth & Frequency   & References \\
             & Code  & Configuration & Channels &  (GHz)    & $\nu$ (MHz) &  \\
\tableline
1992 Jun 27  & C148  & 6A    & 33   &  0.128 &  4786       &\citet{dm}\\
1993 Feb 21  & C148  & 6C    & 33   &  0.128 &  4786, 8640 &\citet{dm}\\
1993 Mar 15  & C148  & 1.5D  & 33   &  0.128 &  4786, 8640 &\citet{dm} \\
1993 Mar 20  & C148  & 1.5B  & 33   &  0.128 &  4786, 8640 &\citet{dm} \\
2015 Jan 01  & CX310 & 6A    & 2049 &  2.048 &  5500, 9000 & Our observation \\
2017 Dec 25  & CX403 & 6C    & 2049 &  2.048 &  5500, 9000 & Our observation \\
2018 Mar 27  & C3229 & EW352 & 2049 &  2.048 &  5500, 9000 & Our observation\\
\tableline
\end{tabular}
\end{table*}

C148 data (pre-CABB) include observations at 4786 and 8640\,MHz over four days in 1992-1993, with four different arrays -- 6A, 6C, 1.5D, and 1.5B respectively. Primary calibration (bandpass and flux density) used source PK\,B1934--638 for all four days, while secondary calibration (phase) used source PKS\,B0407--658 for one day 27$^{th}$ Jun 1992, and PKS\,0407--810 for the other three days. More details can be found in \citet{dm}. CABB data observations were made over three days in 2015-18, with different array configuration (6A, 6C, and EW352) at frequencies of 5500 and 9000\,MHz. Primary calibration used source PKS\,1934--638 (2015 and 2018) and  PKS\,B0823--500 (2017). Source PKS\,B0530--727 was used as a secondary calibration for all CABB data. We reduced the data with the \textsc{miriad}\footnote{http://www.atnf.csiro.au/computing/software/miriad/} \citep{stw} and \textsc{karma}\footnote{http://www.atnf.csiro.au/computing/software/karma/} \citep{go} software packages. 

We used the \textsc{miriad} task \textit{invert} with input parameter, \textit{robust}, which has a range between --2 and 2, where: --2 is uniform weight, and 2 is natural weight. Robust has been selected to be 0 at 5500, 4786 and 9000\,MHz; and 1 at 8640\,MHz. \textsc{miriad} tasks: \textit{mfclean} and \textit{restor} were used to clean and deconvolve CABB data, while the tasks \textit{clean} and \textit{restor} were used to clean and deconvolve pre-CABB data. 

\begin{table}
\small
\caption{Details of \n\ images used in this study\label{tab2}}
\begin{tabular}{@{}cccc@{}}
\tableline
 $\lambda$ & $\nu$ & Beam Size & RMS \\
(cm)       & (MHz) & (arcsec)  & (mJy\,beam$^{-1}$) \\
\tableline
6 & 4786 & 2.7$\times$2.3    & 0.237 \\
6 & 5500 & 1.8$\times$1.3    & 0.031 \\
3 & 8640 & 3.6$\times$2.4    & 0.351 \\
3 & 9000 & 1.10$\times$0.85  & 0.022 \\
\tableline
\end{tabular}
\end{table}

The resulting image properties from this reduction process are summarised  in Table~\ref{tab2}. The image at 4786\,MHz is a combination of all four observing days, while at 8640\,MHz we used only three days because of the strong radio-interference on 27$^{th}$~Jun~1992. This negatively affected the spatial resolution of this image. The images at 5500 and 9000\,MHz combine three days of observations and have the lowest RMS noise due to favourable atmospheric conditions.

\section{Results and Discussion}
 \label{sec:results}

\subsection{Radio morphology}
\begin{figure*}[t]
 \includegraphics[width=0.49\textwidth, trim=0 50 10 60,clip]{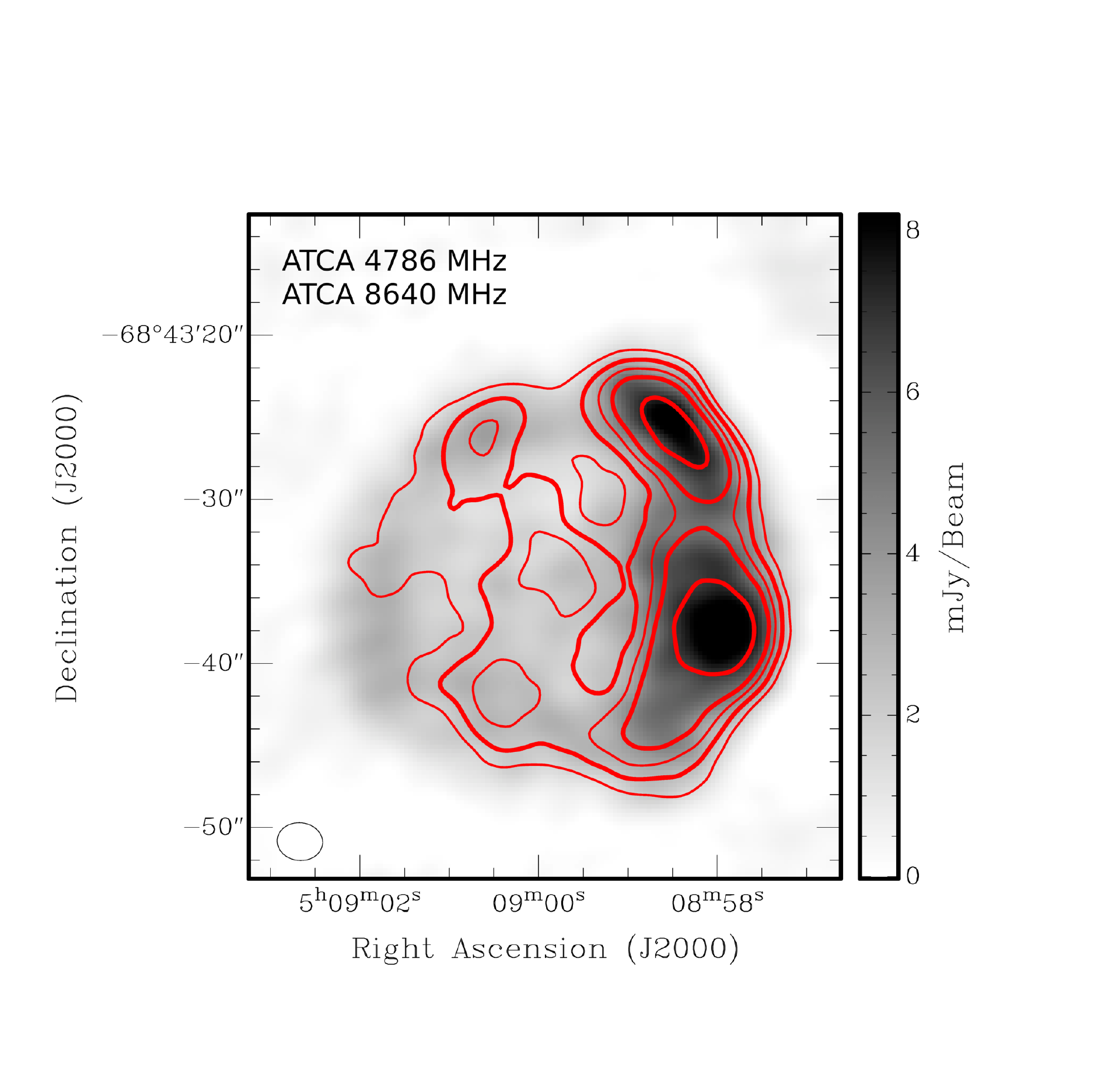}
 \includegraphics[width=0.49\textwidth,trim=0 50 10 60,clip]{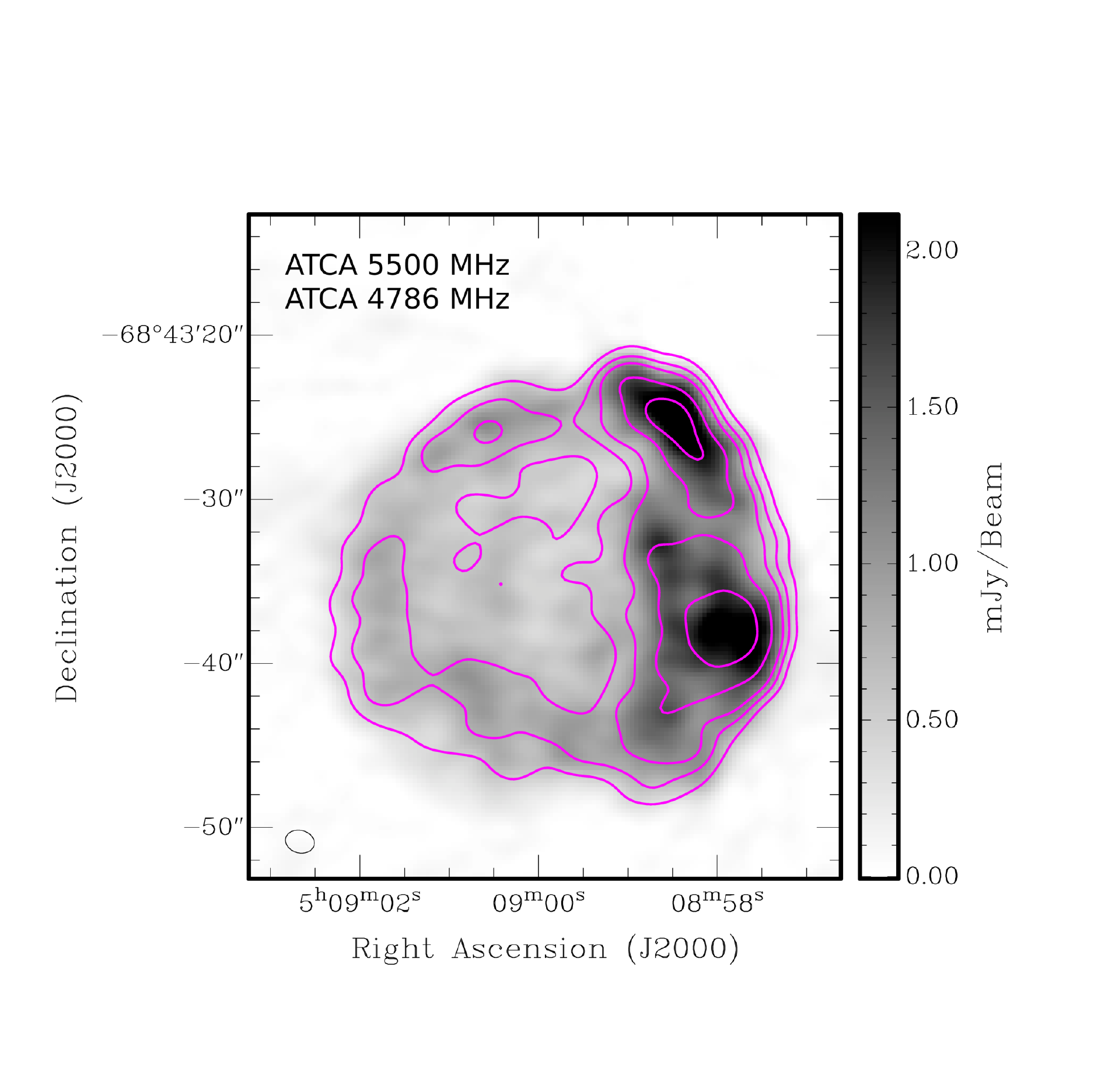} 
 \caption{ATCA (pre-CABB) image of SNR \n~at 4786\,MHz (left) and CABB at 5500\,MHz (right) overlaid with contour map at 8640\,MHz (left; red) and 4786\,MHz (right; magenta). The red contour levels are: 1.5, 2, 2.5, 3, and 4\,mJy\,beam$^{-1}$ while the magenta contour levels are: 1.5, 2.5, 3.5, 5.5, and 7.5\,mJy\,beam$^{-1}$. The ellipse in the lowest left corner is the synthesised beam of 2$\farcs$7$\times$2$\farcs$3 and 1$\farcs$8$\times$1$\farcs$3, respectively. The sidebar on the right hand side of each image represents the used gray scale gradient in mJy\,beam$^{-1}$. The RMS noise of the 4786\,MHz image is 0.24\,mJy\,beam$^{-1}$, 5500\,MHz image is 0.031\,mJy\,beam$^{-1}$, and 0.35\,mJy\,beam$^{-1}$ in the 8640\,MHz image.}
 \label{Fig1}
\end{figure*}

Figure~\ref{Fig1} (\emph{left}) shows the ATCA (pre-CABB) image of SNR \n\ at 4786\,MHz, overlaid with contours of 8640\,MHz emission. The \emph{right} panel shows the ATCA (CABB) image of SNR \n\ at 5500\,MHz, with contours of 4786\,MHz emission.

With a geometric centre of RA(J2000)= 05$^h$08$^m$59.4$^s$ and Dec(J2000)=--68\degr43\arcmin35\arcsec, SNR \n\ shows a somewhat circular shell with significant morphological asymmetries. In particular, there are two bright regions toward the north-west and south-west. As we will discuss further in Section~\ref{sec:comparison}, \n\ is morphologically distinct to other young \citep[$\sim$400\,yrs;][]{2018MNRAS.479.1800R} type\,Ia SNR, which are generally much more symmetric.

\subsection{Radio spectral index} 
 \label{radio}
SNR radio spectra can often be described as a pure power-law of frequency: $S_{\nu}$~$\propto$~$\nu^\alpha$, where $S_{\nu}$ is flux density, $\nu$ is frequency, and $\alpha$ is the spectral index. Such a simple mathematical form stems from the test-particle diffusive shock acceleration (DSA) theory \citep{Axford77, Krymsky77, Bell78, BO78}. To more accurately measure the spectral index of our source, we combine our observations with flux density measurements across a wide range of frequencies from the Murchison Widefield Array \citep[MWA;][]{2018MNRAS.480.2743F}; 84--200\,MHz and Molonglo, Parkes and ATCA (408--14700\,MHz), as shown in Table~\ref{tab3}. 

\begin{table}
\footnotesize
\caption{Flux density measurements of SNR \n\ at multiple frequencies
 \label{tab3}}
\begin{tabular}{@{}cccl@{}}
\tableline
$\nu$ & Flux Density & Telescope & Reference\\
(MHz) &$S_{\nu}$(Jy)\\
\tableline
84&8.02&MWA&\cite{2018MNRAS.480.2743F}\\
107&4.95&MWA&\cite{2018MNRAS.480.2743F}\\
115&4.44&MWA&\cite{2018MNRAS.480.2743F}\\
118& 4.69&MWA& \cite{2018MNRAS.480.2743F}\\
122&4.32&MWA&\cite{2018MNRAS.480.2743F}\\
130&4.19&MWA&\cite{2018MNRAS.480.2743F}\\
143&3.43&MWA&\cite{2018MNRAS.480.2743F}\\
150&3.45&MWA&\cite{2018MNRAS.480.2743F}\\
155&3.42&MWA&\cite{2018MNRAS.480.2743F} \\
158&3.11&MWA&\cite{2018MNRAS.480.2743F}\\
166&2.94&MWA&\cite{2018MNRAS.480.2743F}\\
173&3.10&MWA&\cite{2018MNRAS.480.2743F}\\
189&2.59&MWA&\cite{2018MNRAS.480.2743F}\\
196&2.65&MWA&\cite{2018MNRAS.480.2743F}\\
200&2.80&MWA& \cite{2018MNRAS.480.2743F} \\
408&1.57&Molonglo&\citet{2017ApJS..230....2B}\\
843&0.613&MOST&\citet{2017ApJS..230....2B}\\
1377&0.496&ATCA&\citet{2017ApJS..230....2B}\\
4750&0.429&Parkes&\citet{2017ApJS..230....2B}\\
4800&0.217&ATCA&\citet{2017ApJS..230....2B}\\
4850&0.426&Parkes&\citet{2017ApJS..230....2B}\\
5000&0.360&Parkes&\citet{2017ApJS..230....2B}\\
5500&0.380&ATCA&This work\\
8550&0.226&Parkes&\citet{2017ApJS..230....2B}\\
8640&0.138&ATCA&\citet{2017ApJS..230....2B}\\
9000&0.130&ATCA&This work\\
14700&0.16&Parkes&\citet{2017ApJS..230....2B}\\
\tableline
\end{tabular}
\end{table} 

\begin{figure*}[!htb]
\centering
\centering
\includegraphics[width=0.46\textwidth,trim=0 0 110 0, clip]{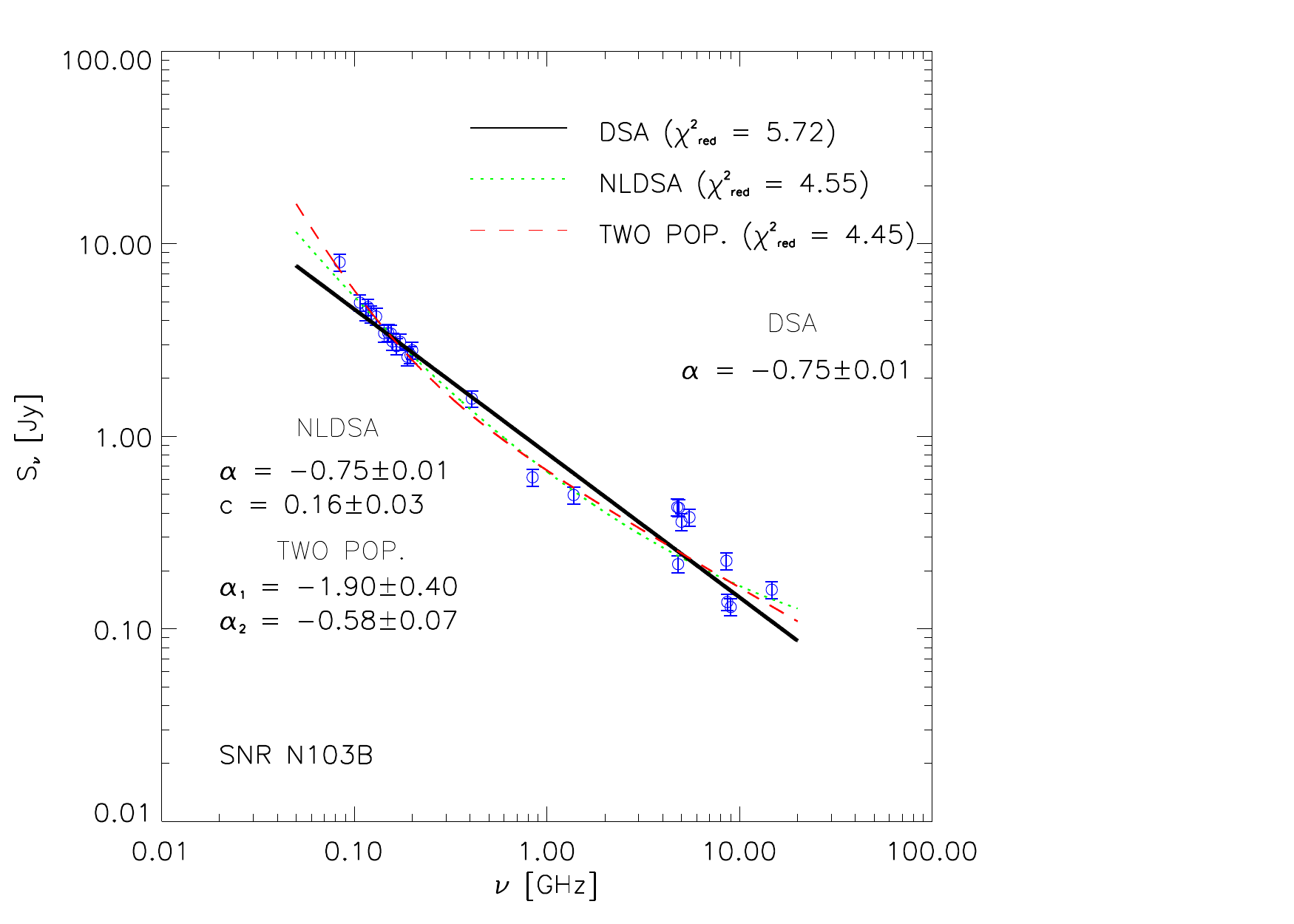}
\includegraphics[width=0.505\textwidth,trim=20 7 12 5,clip]{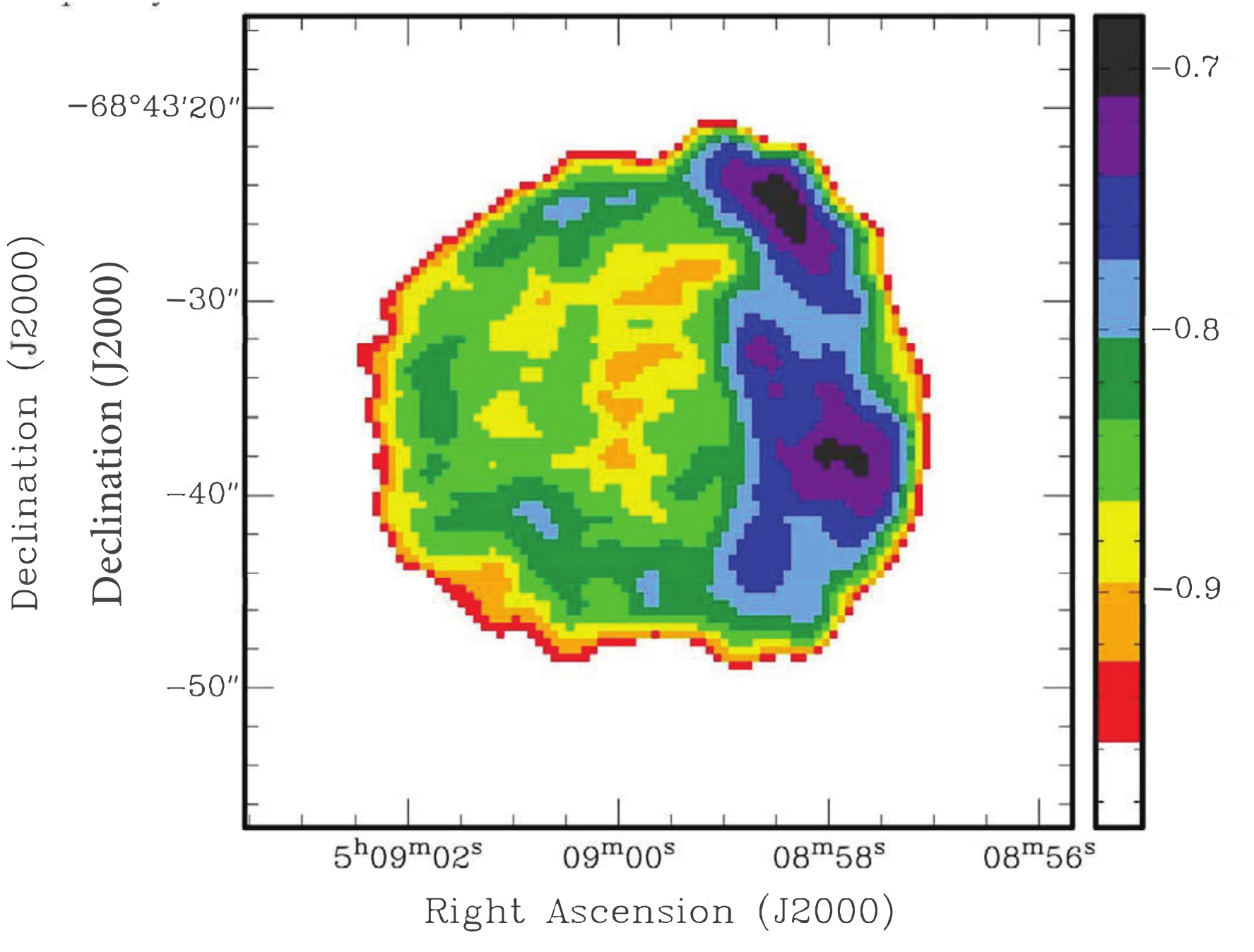} \\
\caption{\emph{Left:} Radio continuum spectrum of SNR \n, from MWA and ATCA data (Table \ref{tab3}). The black solid line shows a linear least squares fit in logarithmic space, giving the spatially integrated spectral index of $\langle\alpha\rangle= -0.75 \pm 0.01$. The simple NLDSA model is represented by a green dotted line, while the model that assumes two different populations of electron is shown as a dashed red line. The relative errors (assumed 10\% uncertainty in each flux density measurement) are used for the error bars on a logarithmic plot. \emph{Right:} Spectral index map for \n\ at 5500\,MHz, showing a shallower index where the emission is brightest. The uncertainty in $\alpha$ is~$<$~0.2.}
\label{fig3}
\end{figure*}

In Figure~\ref{fig3} (\emph{left}) we plot \n\ flux density vs.~frequency. The relative errors (assumed 10\% uncertainty in all flux density measurement) are used for the error bars on a logarithmic plot. The best power-law weighted least-squares fit is shown (thick black line), with the spatially integrated spectral index $\langle\alpha\rangle= -0.75 \pm 0.01$ which is marginally higher, but still consistent with the previous measured value of --0.67 of \citet{dm}. This steeper radio spectral index value is typical of young SNRs \citep{2014Ap&SS.354..541U,2017ApJS..230....2B}. Actually, a couple of theoretical models have been proposed recently to account for such steep spectra observed in a significant number of young SNRs. These include oblique-shock effects \citep{2011MNRAS.418.1208B}, as well as the loss of cosmic-ray energy to turbulence and magnetic field at the non-relativistic quasi-parallel shocks \citep{BMB19}. In addition, \citet{2017MNRAS.468.1616P} demonstrated that the steep overall radio spectral index of SNR G1.9$+$0.3 can be explained only by means of the efficient non-linear diffusive shock acceleration (NLDSA).

Despite the actual scatter in data one can also discuss the apparent curvature in the overall radio spectrum. The curved radio spectra of young SNRs are usually interpreted as a result of an underlying NLDSA process \citep{RE92, Jones03, DELO17}. To model NLDSA effects on the global radio continuum spectrum, we used a simple varying power-law of the form $S_{\nu}\propto\nu^{\alpha+c\log\nu}$, where $c$ represent a curvature parameter. The obvious better fit is obtained (see green dotted line in Figure~\ref{fig3} for NLDSA fit). It has a large curvature ($c=0.16\pm0.03$) in comparison with the value of $0.03$, for the case of Cas A \citep{2015ApJ...805..119O}. It is interesting to note that spatial variations in the spectral index in the case of Cas A remnant have also been proposed to be due to power-laws of different slopes rather than a single curved spectrum \citep{1999ApJ...518..284W}. In that sense, we also tried to interpret the integrated radio continuum of this remnant by assuming synchrotron radiation by two different cosmic-ray electron populations (TWO POP) that possibly reside in the SNR. In fact, we used a simple sum of two power-laws of the form $S_{\nu}=S_{1}\nu^{\alpha_{1}}+S_{2}\nu^{\alpha_{2}}$. The fit is even better in this case but requires a very high value for a steeper spectral index ($\alpha_{1}=-1.90\pm0.40$, $\alpha_{2}=-0.58\pm0.07$), that is not usually seen in SNRs (see red dashed line in Figure~\ref{fig3} for TWO POP.~fit). Of course, a high scatter in data points especially between 4--5\,GHz prevent us from making firm conclusions on the particle acceleration properties based solely on the integrated radio continuum. For all the weighted least-squares fitting of the radio continuum we have used the \verb"MPFIT"\footnote{http://purl.com/net/mpfit} \citep{MPFIT} package written in \verb"IDL", with starting values estimated from the data.

On the other hand, the analysis of spatial variations in the radio continuum spectral index within the SNR is very important \citep[see e.g.~discussion on the properties of SNRs Tycho, Cas A, as well as SN 1978A; ][]{2000ApJ...529..453K, 1996ApJ...456..234A, 2013ApJ...767...98Z}. Such spatial variations are generally linked to the electron acceleration processes that among other processes depend on the evolution of the SNR. It can be concluded that the synchrotron radio spectral index traces the distribution of energy among cosmic-ray electron populations and can probe the compression ratio of the shock on a local scale. That is why we can use these results to study properties of the particle acceleration mechanisms. In that sense, understanding of the spatial variation of the spectral index and emission across young SNRs is imperative.

Figure~\ref{fig3} (\emph{right}) shows a map of the spectral index for \n. It can be seen that the spectral index distribution across the remnant is not at all uniform. The spectral index is shallower (between --0.7 an --0.8) in the bright regions on the Western side of \n, and steeper in the centre and Eastern side of the SNR (from --0.8 to around --0.9\textbf{; $0.1<\Delta_{\alpha}<0.2$}). This could indicate different emission and/or particle acceleration scenarios for different regions (various population of electrons are contributing). It should be noted that the steep spectral indices are also observed in the regions where SNR interact with molecular environment as seen in Figure~\ref{XrayFig1}, contrary to the usual expectations that involve simple compression (amplification) of magnetic field. This is somewhat similar to what we observed in the case of Galactic SNR Vela Jr. \citep{2018ApJ...866...76M} and LMC SNR\,N\,63A \citep{2019ApJ...873...40S}. However, both of these SNRs are most likely result of CC explosion. If the electron energy spectrum hardens at lower frequencies, regions in which the magnetic field is higher than the surroundings will both appear brighter and have a flatter spectrum than the surroundings. Actually, regions in which the magnetic field is stronger would appear both brighter and with a flatter spectrum. Of course, if the shock wave is significantly decelerated due to the interaction with the molecular cloud environment, we expect less efficient particle acceleration which leads to the steeper radio spectral indices. In addition, optical images show H$\alpha$ emission along the entire periphery of the Western portion of the shock, with [\mbox{O\,{\sc iii}}] and [\mbox{S\,{\sc ii}}] lines emitted from a few dense clumps where the shock has probably become radiative \citep{2014ApJ...790..139W}. These regions coincide with spectral indices between --0.7 and --0.8. 

An important clue for this analysis can be found in the linear polarisation maps at 5500 and 9000\,MHz discussed in Section~\ref{pol} (see also Figure~\ref{Fig4}). Linear polarisation is detected mainly where the emission is brightest, in the Western part of the remnant. However, a bit less pronounced linear polarisation is also seen in the Eastern parts of the SNR. Radial and oblique orientation of the ordered magnetic field is more dominant in the regions of high radio and X-ray emission (Western parts). Quasi-perpendicular shock geometries are more pronounced in the Eastern parts of the SNR, which can explain less efficient particle acceleration in these regions. Still, we note that \citet{2018JPlPh..84c7101C} have showed that the high-energy particles can be effectively reflected and accelerated regardless of shock inclination via so-called diffusive shock re-acceleration process. They found that re-accelerated Galactic cosmic-rays can drive the streaming instability in the shock upstream and produce effective magnetic field amplification. This can then trigger the injection of thermal particles even at quasi-perpendicular shocks.

Finally, a word of caution regarding this discussion. As the estimated uncertainty for the measured spectral indices is between 0.1--0.2, it is therefore not possible to draw any firm conclusions about this issue. Especially, we note that the steepest $\alpha$ regions are correlated with the lowest signal to noise regions.  

\subsection{Expansion}
There are many difficulties measuring the expansion rate for the SNR shell, because the CABB and pre-CABB data were taken using different instrument setups and different band-passes. Therefore, differences in SNR shell size at different epochs (about 26 years apart) are difficult to disentangle from the above mention observational and instrumental effects.

Nevertheless, we can place limits on the expansion velocity of the shell. We compare the pre-CABB data with the CABB image data, smoothing the lower spatial resolution pre CABB image data (4786\,MHz) to the higher resolution of the CABB data (5500\,MHz) for an equivalent full width half maximum (FWHM) of 0.5--1\,arcsec resolution. 
If we interpret the resulting differences in our radio images as due to motion (rather than due to brightening, intrinsic or instrumental), then our estimate of the expansion velocity is in range of 4000--9000\,km\,s$^{-1}$, this value is consistent with the result of \citet{2041-8205-865-2-L13} $4170^{+1280}_{-1310}$~\kms  ($\sim$0.017\,arcsec\,yr$^{-1}$), which are based on Chandra X-ray observations.


\subsection{Polarisation} 
 \label{pol}

\begin{figure*}[htb!]
\centering
 
\hspace{-0.5\textwidth}a)\hspace{0.47\textwidth}b) \\
\includegraphics[width=0.48\textwidth,trim=0 13 0 15,clip]{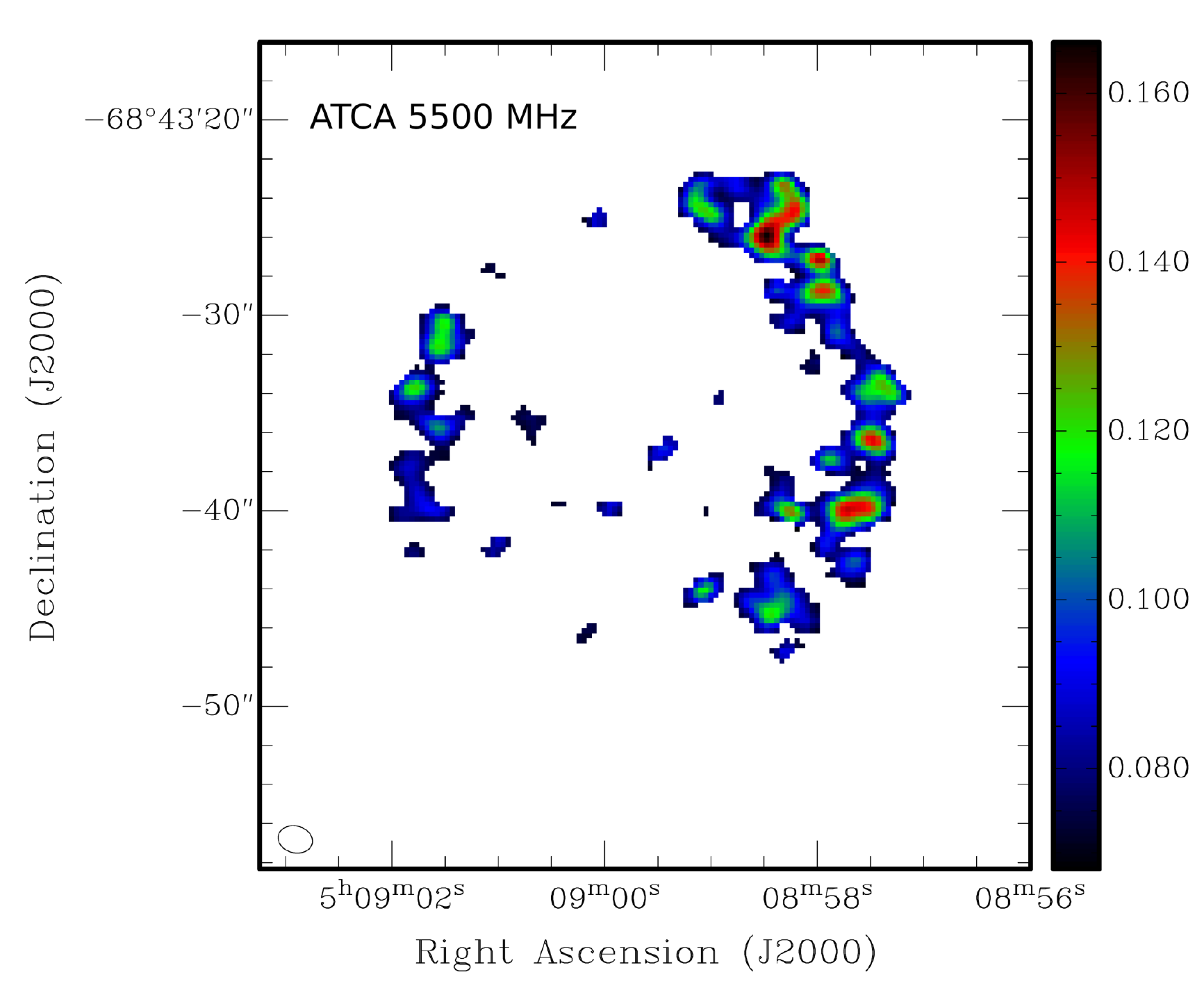}
\includegraphics[width=0.51\textwidth,trim=0 0 0 0,clip]{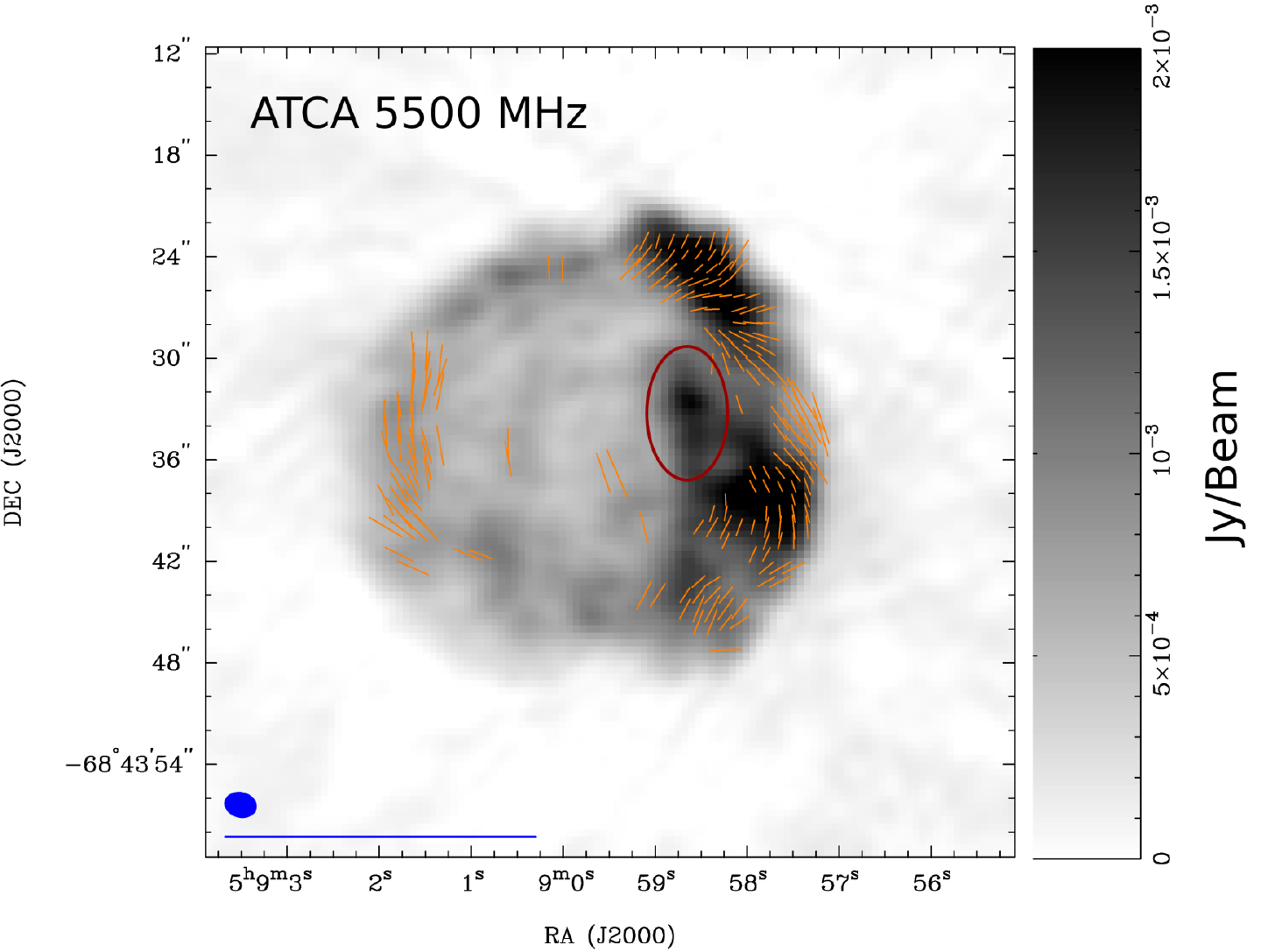}\\
\hspace{-0.5\textwidth}c)\hspace{0.47\textwidth}d) \\
\includegraphics[width=0.48\textwidth,trim=0 10 0 15,clip]{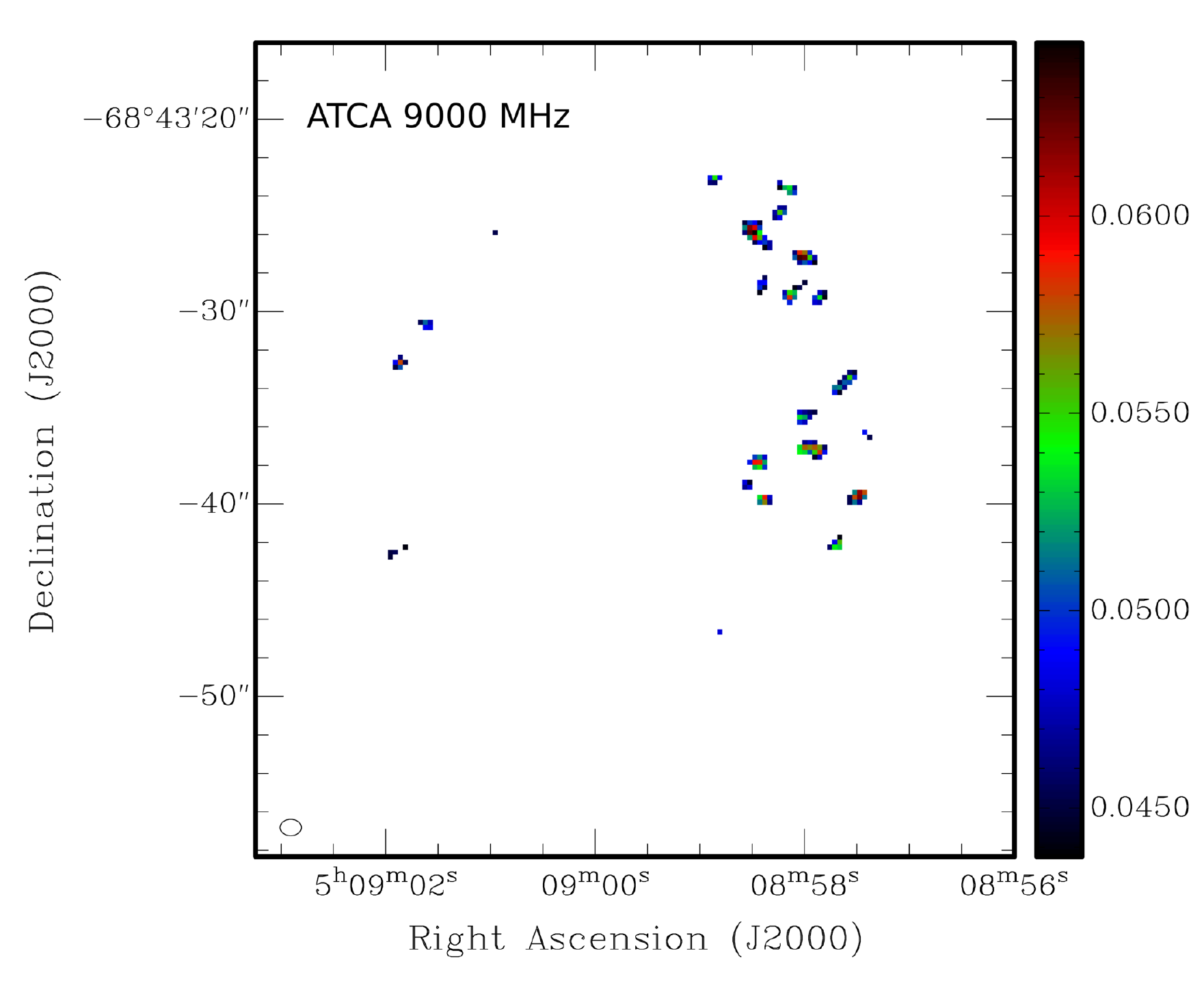} 
\includegraphics[width=0.51\textwidth,clip]{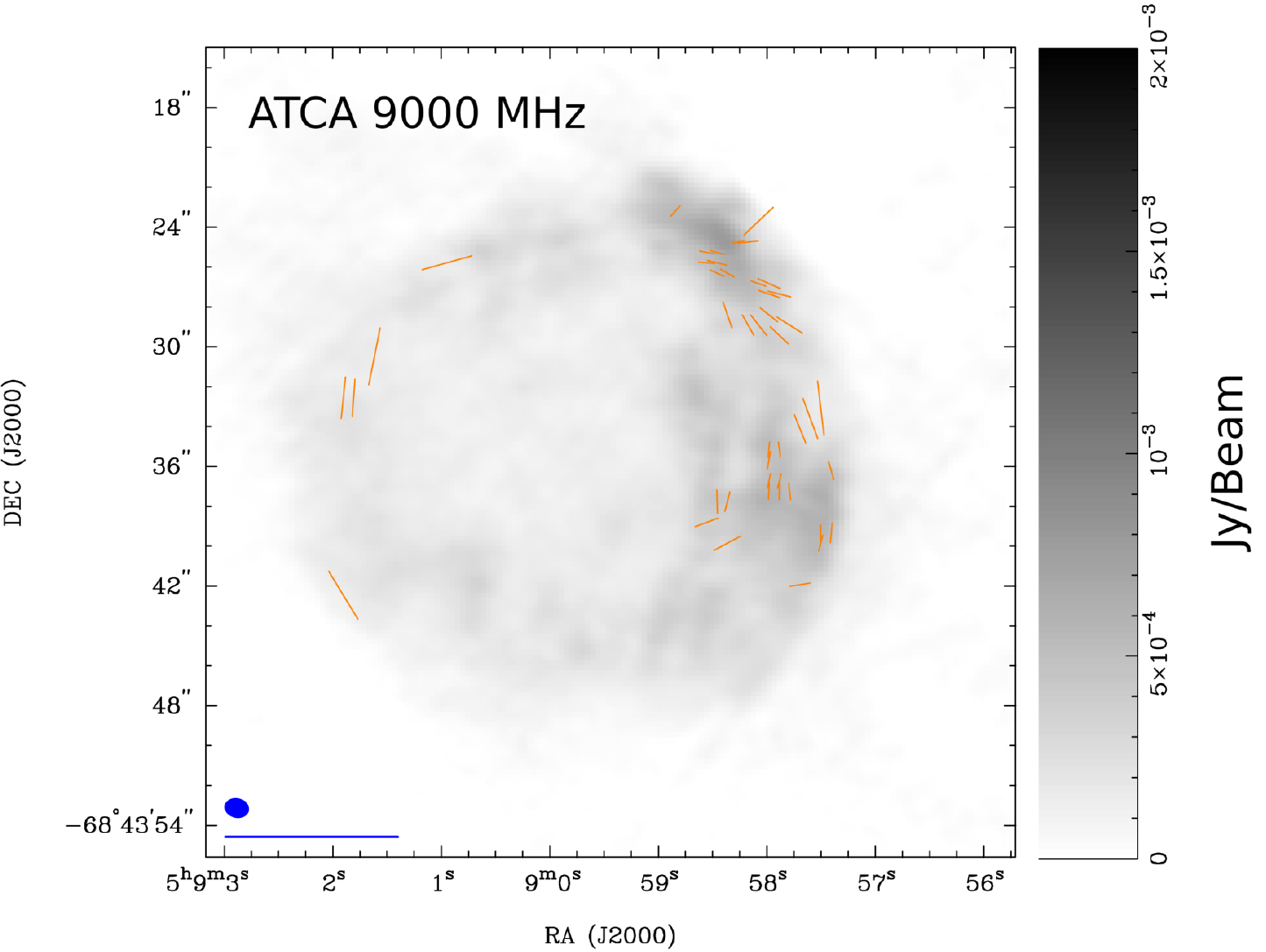} \\
\caption{a) Polarisation intensity map of SNR \n\ at 5500\,MHz. The ellipse in the lower left corner shows the synthesised beam of 1\farcs8$\times$1\farcs3. b) Fractional Polarisation vectors (orange) overlaid on top of the intensity map (gray scale) of SNR \n\ at 5500\,MHz. The blue ellipse is similar to (a) and the blue line below the ellipse represents 100\% polarisation. Interestingly, we note a radio thermal emission toward the west (red ellipse). c) Polarisation intensity map of SNR \n\ at 9000\,MHz. The ellipse in the lower left corner shows the synthesised beam of 1\farcs1$\times$0\farcs85. d) Fractional Polarisation vectors (orange) overlaid on top of the intensity map (gray scale) of SNR \n\ at 9000\,MHz. The blue ellipse is similar to (c) and the blue line below the ellipse represents 100\% polarisation.
\label{Fig4}}
\end{figure*}


 


Polarisation is calculated using the Stokes parameters:
\begin{equation} \label{eq1}
P=\frac{\sqrt{S^2_{Q}+S^2_{U}}}{S_{I}} ~,
\end{equation}
where $P$ is the mean fractional polarisation, $S_{Q}$, $S_{U}$, and $S_{I}$ are integrated intensities for the $Q$, $U$, and $I$ Stokes parameters. The overall fractional polarisation of \n\ is 8$\pm$1\% at 5500\,MHz and 13$\pm$3\% at 9000\,MHz. This is somewhat higher than the value of 5\% determined by \citet{dm} at 4786\,MHz. However, we emphasise that the ATCA polarisation capabilities back in 1992--1993 were significantly lower. Also, the depolarisation effect might play a role caused by differential Faraday rotation.

Figure \ref{Fig4} shows polarisation intensity map at 5500\,MHz (\ref{Fig4}a), fractional polarisation map at 5500\,MHz (\ref{Fig4}b), polarisation intensity map at 9000\,MHz (\ref{Fig4}c), and fractional polarisation map at 9000\,MHz (\ref{Fig4}d). All polarisation images haven't been corrected for Faraday rotation. The average polarisation intensity at 5500 and 9000\,MHz are $\sim$0.09 and $\sim$0.05, respectively. The fractional polarisation map at 5500\,MHz reveals linear polarisation, particularly where the emission is brightest in the east and north-west of the remnant. This probably indicates the existence of coherent magnetic fields where the shell impacts the ISM. The 9000\,MHz images show lower levels of polarisation, but a similar pattern of intensity and direction with respect to the radio emission.

Note an exception: the bright structure towards the west (marked with red ellipse) is (total intensity) bright but unpolarised. This structure is obvious in our new CABB images but unclear in pre-CABB images. This is due to the fact that our new CABB images have better resolution and lower RMS noise. Another possibility, however less likely, is that this region either brightened or expanded significantly faster than the rest of this remnant. The lack of polarisation may indicate intense radio thermal emission coming from this region.

\subsection{Faraday Rotation and Magnetic Field}
We can measure interstellar magnetic fields through their effect on the propagation of linearly polarized radiation. As the polarized emission from a radio source passes through a magnetized, ionized plasma, the plane of polarisation rotated due to the different phase velocities of the two polarisation modes. This Faraday rotation changes the position angle of the emission by an amount which depends on the rotation measure (RM):
\begin{equation}
RM=811.9\int_{0}^{L}n_{e}B_{\|}dl ~,
\label{eq2}    
\end{equation}
where RM is in rad\,m$^{-2}$, $n_{e}$ is the electron density in $cm^{-3}$, $B_{\|}$ is the line of sight magnetic field strength in $\mu$G, and $L$ is the path length through the Faraday rotating medium in kpc. The change in position angle of the radiation is:
\begin{equation}
\Delta_{X}= X - X_0 = RM~\lambda^{2} ~,
\end{equation}
Where $X$ is the observed position angle at wavelength $\lambda$, and $X_0$ is the intrinsic position angle of the polarized emission. To overcome the $n~\pi$ ambiguity in measurements of the position angle, we require observations at three or more wavelengths \citep{cl}. To calculate RM, we split the 2048\,MHz bandwidth at 5500\,MHz into three 680\,MHz sub bands (4817, 5500, and 6183\,MHz).

\begin{figure*}[ht!]
\includegraphics[width=\textwidth]{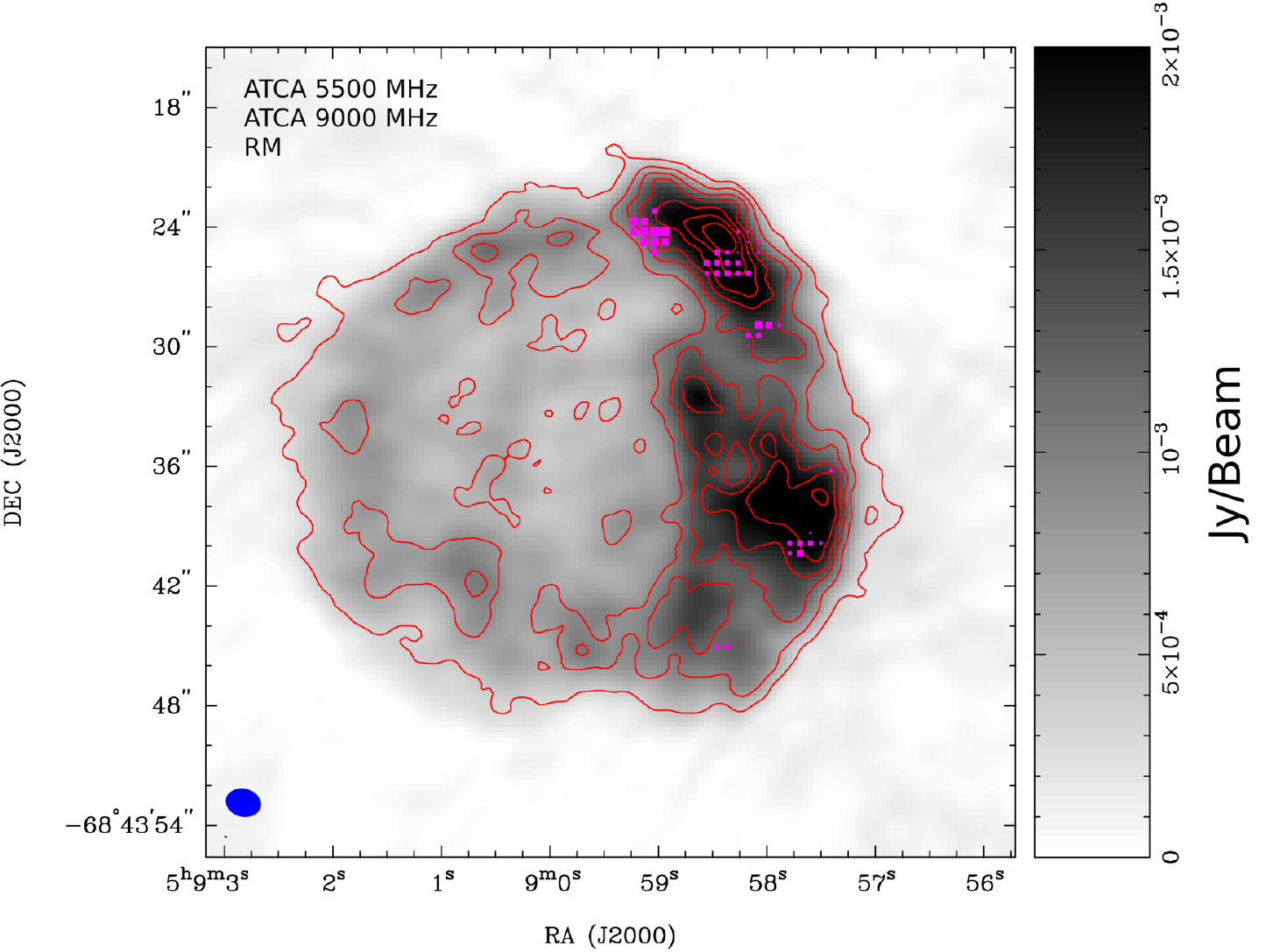}\\
\caption{SNR \n\ at 5500\,MHz (gray scale) overlaid with 9000\,MHz contours. The contour levels are: 0.1, 0.2, 0.3, 0.4, 0.5, 0.6 and 0.7\,mJy\,beam$^{-1}$. The pink boxes represent RM estimated from position angles associated with linear polarisation. All boxes have positive values, the maximum value of RM is $\approx$460\,rad\,m$^{-2}$. The blue ellipse in the lower left corner represents the synthesised beam of 1\farcs8$\times$1\farcs3\label{Fig6}}
\end{figure*}

Figure~\ref{Fig6} shows the measured RM for \n. The mean value for RM in this SNR is 200$\pm$20\,rad\,m$^{-2}$ (with a peak of $\approx$460\,rad\,m$^{-2}$). Similar to the polarisation, the greatest rotation measure can be seen close to where the total intensity emission is brightest.

Having measured $RM$, we can use Equation~\ref{eq2} to estimate the magnetic field strength. Table~\ref{tab4} summarizes different estimates of the electron density of \n. \citet{li} estimated the mean electron number density toward the optical knots or post-shock gas of \n\ to be $\sim$500--5000\,cm$^{-3}$.  \cite{Wi} calculated the post-shock gas's electron number density to be $\sim$45\,cm$^{-3}$ using IR data. \citet{So} estimated the plasma electron number density to be $\sim$10--47\,cm$^{-3}$, while the estimate by \citet{Van} is $\sim$7--25\,cm$^{-3}$ and \cite{So} reported an electron number density of $\sim$2--10\,cm$^{-3}$ in the ambient gas (pre-shock density). Therefore, the Faraday rotation is through the shocked ionised gas, and thus the n$_e$ relevant is 4 times that value. That would makes it $\sim$8--40~cm$^3$, so closer to \citet{Van} or \citet{Wi} estimate, and narrow down the range of magnetic field. There is no doubt that the range of n$_e$ shows wide varieties but some are less applicable than others. The one for the optical knots (high n$_e$) are for small clumps and not likely representing the overall medium density. 

Using the average value of RM (200\,rad\,m$^{-2}$), and $L$ as the thickness of the compressed shell of the SNR ($\sim$1.5\,pc), we find that the magnetic field strength values for \n\ vary from 0.03\,$\mu$G to 82.1\,$\mu$G, with most values clustered around 16.4\,$\mu$G. The specific values are shown in Table~\ref{tab4}. 

We also use the equipartition formulae\footnote{http://poincare.matf.bg.ac.rs/$\sim$arbo/eqp/} \citep{Arbutina_2012,2013ApJ...777...31A,Uro_evi__2018} to estimate the magnetic field strength for this SNR. This derivation is purely analytical, accommodated especially for the estimation of the magnetic field strength in SNRs. The average equipartition field over the whole shell of \n\ is $\sim$235\,$\mu$G with an estimated minimum energy\footnote{We use the following \n\ values: $\theta$=0.24\arcmin, $\kappa=0$, S$_{\rm 1\,GHz}$=0.903 and f=0.25; for $\kappa\neq0$ we estimate the average equipartition field of 396\,$\mu$G with an estimated minimum energy of E$_{\rm min}$=1.8$\times10^{49}$\,erg which is probable overestimate, because physical background gives better equipartition arguments for $\kappa=0$ \citep{Uro_evi__2018}.} of E$_{\rm min}$=6.3$\times10^{48}$\,erg. This value is typical of young SNRs with a strongly amplified magnetic field. For example, its cousin and neighbouring LMC SNR\,J0509--6731 has an estimated average equipartition magnetic field strength of $\sim$168\,$\mu$G \citep{2014MNRAS.440.3220B}, while the much older (20--25\,kyr) type\,Ia MCSNR\,J0508--6902 has a magnetic field strength of $\sim$28\,$\mu$G \citep{2014MNRAS.439.1110B} or the middle-aged MCSNR\,J0530--7007 where \cite{2012A&A...540A..25D} estimated the equipartition magnetic field over the whole shell to be $\sim$53\,$\mu$G. 



\begin{table}
\small
\caption{Electron density and magnetic field strength for \n\ based on a various estimates for the electron density
\label{tab4} 
}
\begin{tabular}{@{}clcc@{}}
\tableline
$n_{e}$ & References  & B\\
(cm$^{-3}$)&~ & ($\mu$G)\\
\tableline
500--5000   &\cite{li}& 0.3--0.03  \\
45  &   \cite{Wi} &3.6 &\\
10--47 &   \cite{So} &16.4--3.4 \\
7--25  &   \cite{Van} &23.5--6.6  \\
2--10&   \cite{So}&82.1--16.4 \\
\tableline
\end{tabular}
\end{table}

\subsection{H$\alpha$ and X-ray Emission}
\begin{figure*}[ht!]
 \centering
\includegraphics[width=\textwidth]{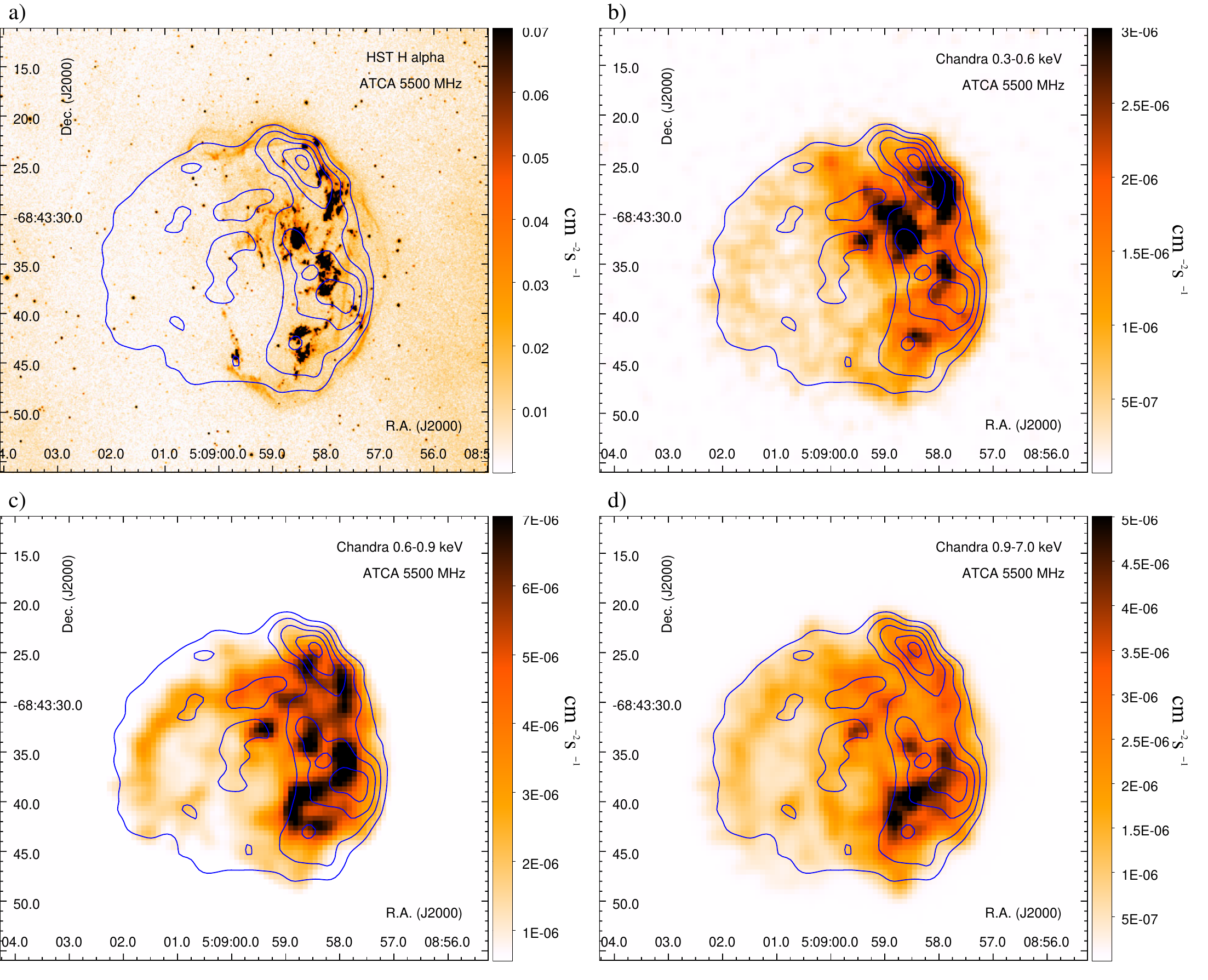} 
 \caption{a) Hubble Space Telescope H$\alpha$ image of \n~\citep{li}. b), c) and d) Chandra 0.3--0.6, 0.6--0.9 and 0.9--7.0\,keV images of \n~\citep{Lewis:2003}. All images have ATCA 5500\,MHz radio continuum contours (this study) overlaid. }
 \label{XrayWrad}
\end{figure*}

In Figure~\ref{XrayWrad} we compare our new ATCA 5500\,MHz radio continuum observations to Hubble Space Telescope H$\alpha$ emission \citep{li}, and Chandra `soft', `medium' and `hard' X-ray emission (0.3--0.6, 0.6--0.9 and 0.9--7.0\,keV, respectively). \citet{li} have previously shown thermal emission in H$\alpha$ that corresponds to soft X-ray emission seen by \citet{Lewis:2003}, and is particularly prominent towards regions with dense optical knots. These thermal signatures of an ionised circumstellar medium are not expected to correspond to the \n\ radio emission which traces a non-thermal component from electrons accelerated by the SNR shock. 

Nevertheless, some correlation between radio continuum and hard non-thermal X-rays is expected. As seen in Figure\,\ref{XrayWrad}d (0.9--7.0\,keV), the hard X-ray emission follows the radio emission, but not particularly closely. Although the large-scale structure of \n\ radio continuum emission is evidently enhanced in the Western part of SNR, comparing with other two frequency (optical and X-ray) emissions no exact/precise matching can be seen. There are many lines (Ne, Mg, Si, S, Ar, even Fe K) in 0.9--7.0\,keV band, which actually most likely dominates the emission from \n. Perhaps, this is the main reason why even in that band they are no particularly striking radio-to-X-ray match.

\section{Comparison to Similar Age and Type\,Ia SNRs} 
 \label{sec:comparison}

\begin{table*}[ht!]
\footnotesize
\caption{Young ($\leq$1000\,yrs) type\,Ia SNRs in Galaxy (MW) and LMC. The proposed degeneracy scenario listed in Column\,8 are abbreviated as SD for single degenerate and DD for double degenerate type\,Ia explosion. No firm evidence of surviving companions have been found in any of these SNRs. We note that different methods were used to estimate of the SNRs magnetic field. 
 \label{tab0} }
\begin{tabular}{@{}llccccccl@{}} 
\tableline
SNR  & Host  & Age  & Diameter  & Distance   &Spectral Index  &Avg. Polarisation  &Mag. Field  &Deg.    \\
Name         & Galaxy            &(year) &(pc)    &(kpc)        & ($\alpha$)             &     (\%)   &($\mu$G)       &Type\\
\tableline 
Tycho      & MW   & 425$^{1}$    & 3.5--7.2$^{1}$     &   1.5--3.1$^{1}$   &  --0.58$^{2}$ &  20--30 at 4872\,MHz$^{3}$&50--400$^{4}$ & DD$^{5}$   \\
SN1006     & MW   & 1000$^{6}$   & $\sim$19$^{6}$      &   $\sim$2.18$^{6}$    & --0.6$^{2}$ & $\sim$17 at 1400\,MHz$^{7}$&30--40$^{8}$ &  DD$^{9}$ \\
Kepler     & MW   & 407$^{10}$    & $\sim$8.18$^{10}$    &   $\sim$7$^{10}$   &  --0.64$^{2}$  & $\sim$6 at 4835\,MHz$^{11}$ &$\sim$414$^{12}$ & SD$^{13}$ \\
G1.9+0.3   & MW   & $\sim$120$^{14}$         & 4$^{15}$ &   8.5$^{15}$  &  --0.81$^{16}$  & 6 at 5500\,MHz$^{17}$ &273$^{17}$ & DD$^{18}$\\ 
J0509--6731 & LMC         & $\sim$310$^{19}$    & 7.4$^{20}$    &   50      &  $\sim$--0.73$^{20}$  & $\sim$26 at 5500\,MHz$^{20}$& 168$^{20}$ & DD$^{21}$ \\
LHG\,26    & LMC         & $\sim$600$^{22}$   & 8.3$^{22}$    &   50      &  $\sim$--0.54$^{23}$ & $\sim$8 at 5500\,MHz$^{23}$&$\sim$171$^{23}$ & DD$^{24}$ \\
 \n        & LMC         & 380--860$^{25}$    & 6.8$^{26}$    &   50      & $\sim$--0.75 & $\sim$ 8 at 5500\,MHz&0.03--82.1 & SD$^{27}$ \\
\tableline
\end{tabular}
    \begin{tablenotes}
      \footnotesize
      \item References: (1) \cite{1538-4357-545-1-L53}, (2) \cite{2014BASI...42...47G}, (3) \cite{1991AJ....101.2151D}, (4) \cite{2015ApJ...812..101T}, (5) \cite{2017NatAs...1..800W}, (6) \cite{0004-637X-585-1-324}, (7) \cite{2013AJ....145..104R}, (8) \cite{2006ESASP.604..319V}, (9) \cite{2018MNRAS.479..192K}, (10) \cite{2012ApJ...756....6P}, (11)  \cite{2002ApJ...580..914D}, (12) \cite{2012ApJ...746...79A}, (13) \cite{2019MNRAS.482.5651M}, (14) \cite{2017MNRAS.468.1616P}, (15) \cite{2008ApJ...680L..41R} (16) Luken et al. (2019; in prep), (17) \cite{2014SerAJ.189...41D} (18) \cite{2013ApJ...771L...9B}, (19) \cite{2018MNRAS.479.1800R}, (20) \cite{2014MNRAS.440.3220B}, (21) \cite{2012Natur.481..164S}, (22) \cite{1538-4357-642-2-L141}, (23) \cite{2012SerAJ.185...25B}, (24) \cite{2012ApJ...747L..19E}, (25) \cite{res}, (26) \cite{2017ApJS..230....2B}, and (27) \cite{gh}.
    \end{tablenotes}
\end{table*}

We compare the morphological and physical characteristics of young type\,Ia SNRs including their symmetry, average fractional polarisation, alignment between X-ray, optical \& radio emission distribution to degeneracy models as listed in Table~\ref{tab0}. Although, no firm evidences for surviving progenitor companions have been found in any of the proposed young type\,Ia SNRs, we 'predict' the most likely degeneracy scenarios based on common characteristics as found in the literature and respective assumptions therein\footnote{We also emphasise that most of these young SNRs have no firm type\,Ia classification either.}. Table \ref{tab0} shows young (age~$\leq$~1000\,years) type\,Ia SNRs in the Milky Way and the LMC.  

\paragraph{Diameter:} The SNRs from our sample show reasonably similar sizes, which most likely reflects our selection criteria of being young age ($\leq$1000\,yrs). Using the usual linear (or slightly less than linear) expansion and assuming a canonical SN energy of $10^{51}$ ergs, the spread in diameter could be more reflecting different ages and explosion energy.



\paragraph{Spectral Index:} The spectral index ($\alpha$) is similar for the SNRs, with a slight preference for older SNR having shallower spectral indices. In particular, \n\ is very similar to J0509--6731, with $\alpha = -0.73 \pm 0.02$ \citep{2014MNRAS.440.3220B}.

Similarly, the radio spectral index among young type\,Ia SNRs (see Table\ref{tab0}) varies from --0.81 in the youngest Galactic SNR G1.9+0.3 to --0.6 in the oldest (SN1006). As expected, the steepness of the radio spectral index is proportional to their still young age, but we don't see any obvious connection between the proposed degeneracy model and spectral index.

\paragraph{Morphology:} The Galactic environment plays an important role in the morphological characteristics of an evolving SNR. Intriguingly, the six other young type\,Ia SNRs from this sample are significantly more circular and symmetric than \n. \citet{2011ApJ...732..114L} suggested a link between the spherical thermal X-ray morphology and the remnant type, where those SNRs resulting from a type\,Ia SN explosion were more spherical than those from CC supernovae. However, we shouldn't forget that at such young age of their evolution ($<$1000\,yrs) any SNR would keep its circular shape to some degree \citep{2017ApJS..230....2B}. 

J0509--6731, exhibits an almost circular (`doughnut') morphology of 7.4\,pc, with brightened regions towards the south-western limb and a second brightened inner ring which is only seen in the radio continuum \citep{2014MNRAS.440.3220B,2018MNRAS.479.1800R}. We see a very similar morphology in all other young type\,Ia SNRs from our sample except \n, which shows an obvious asymmetric distribution as described earlier. 

We note that J0509--6731 and \n, are positioned $\sim$1.5\degr and $\sim$0.5\degr respectively from the LMC's main optical bar. We might expect the expansion rate of \n\ to be less than that of J0509--6731, since it is located closer to the mid-plane of the LMC where the density of cold gas and dust is typically higher than in the outer regions of dwarf irregular galaxies. This cold gas may be the result of in-fall but due to secular evolutionary processes eventually compressed to form stars and at time-scales ($\sim$ Gyr) due to Galactic disk instabilities form bar and spiral structure typically found in many disk galaxies. This in turn could also explain the fact that the outer radio emission of \n\ is asymmetric and the lower fractional polarisation found around \n\ as compared to J0509--6731 since wind speeds would become truncated. 

The angular size of type\,Ia SNRs at this critical stage of evolution shows interesting tendencies. Apart from the oldest (SN1006) and youngest (G1.9+0.3) SNRs from our sample, all five other SNRs have the same size of 7--8\,pc. Morphologically, we can see the evidence that \n\ as a type\,Ia SD explosion is unique (among this sample) for its asymmetric appearance. A plausible explanation, assuming the SN explosion is of the SD type, is that the nearby giant companion star may have preferentially `blocked' the explosion front from freely propagating in one direction, while in the other having evolved relatively `unobstructed' into the ISM. On average, this would make DD expansion velocities significantly larger than they would be for SD events with apparent radii somewhat larger for a fixed age. However, most hydro-dynamical simulations of SN type\,Ia shock propagating show that a gap from the presence of the companion is quickly `forgotten'. It's more likely that the asymmetrical structure of \n\ comes from the larger scale environment (possibly affected by the progenitor system winds).

\paragraph{Polarisation and Magnetic Field:} The expectation is that the type\,Ia SNR is expanding into a homogeneous regular magnetic field and it is evolved in such a way that the ambient medium dominates the expansion. The RM around the shell should be the same, with a gradient along the shell depending on the angle between the ambient magnetic field and the plane of the sky. In the case of a very young SNR, the magnetic field should be radial in 3-D space, in which case RM symmetry should be observed. However, in our case the magnetic field is concentrated in the Western side of the remnant (figure~\ref{Fig4}). Therefore, we notice that the RM is distributed in the west-side of the remnant only (figure~\ref{Fig6}).

\cite{2014MNRAS.440.3220B} estimated the $P$ value for J0509--6731 at 5500\,MHz of 26$\pm$13\%, which is somewhat higher than the value for \n\ ($\sim$8\% at 5500\,MHz). This may be weakly correlated to age, since J0509--6731 is a younger SNR (almost half the age of \n). However, the same fractional polarisation value has been observed in another young LMC SNR -- LHG\,26 -- both remnants are of approximately the same age (Table~\ref{tab0}). On the other hand, the young Galactic SNRs show a disparity in $P$ values. The $P$ value for the youngest Galactic SNR (G1.9+0.3) is very low in comparison to Tycho and SN\,1006 (see Table~\ref{tab0}). While the diversity in $P$ values might be due to the difference in the ambient density and age, it also may indicate different types of SNe. Specifically, Luken et al. (2019; in prep) show that the alleged type\,Ia SNR G1.9+0.3 might actually be the product of a core collapse SN based on RM, shape and strength of its polarisation vectors.

Previous estimates of magnetic field strength in \n\ vary between 0.03--82.1\,$\mu$G (Table~\ref{tab4}). Yet a stronger equipatrition field of 168\,$\mu$G was estimated for J0509--6731 by \cite{2014MNRAS.440.3220B}, which is still lower than our estimate of \n\ $\sim$235\,$\mu$G. A similar situation can be seen in other type\,Ia SNRs (Table~\ref{tab0}). Although this SNR is a relatively young one, the equipartition assumption is not so proper for the determination of the magnetic field strength (it can be estimated only to an order of magnitude \citep[see][]{Uro_evi__2018}. In any case, the estimated value of 168\,$\mu$G can be explained by the magnetic field amplification at the strong shock of the relatively young and highly luminous SNR. Based on the young type Ia\,SNRs in this study, we can not see any clear differences in polarisation nor magnetic field strength that would lead us to conclude a distinct difference between single or double degenerate SN scenarios. However, further studies with larger sample sizes and better constrained parameters are required to conclusively rule out any connection. 
\paragraph{The surface brightness to diameter diagram:} The position of LMC SNR \n\ in the surface brightness to diameter ($\Sigma$--\textit{D}) diagram ($\Sigma$=6$\times10^{-19}$\,W\,m$^{-2}$\,Hz$^{-1}$\,sr$^{-1}$, D=6.8\,pc) by \cite{2018ApJ...852...84P}, suggests that this remnant is in the early Sedov phase, with an explosion energy of 1--1.5$\times10^{51}$\,erg, which evolves in an environment with a density of 0.02--0.2\,cm$^{-3}$. These values are as expected if \n\ is interacting with a molecular cloud \citep{2018ApJ...867....7S} where the average density is somewhat higher compared to the rest of type\,Ia SNRs.

This result is also in a good accordance with the observed steep overall radio spectral index. Actually, particle acceleration is most efficient exactly in the early stage of the Sedov-Taylor phase. In fact, in this evolutionary stage, the particle acceleration is more efficient than in the ejecta dominated free expansion phase.

Our estimate of \n\ surface brightness ($\Sigma$) is comparable to values found for Galactic remnants in rarefied environments, such as LMC SNR\,J0509--6731 ($\Sigma$= 1.1$\times10^{-19}$\,W\,m$^{-2}$\,Hz$^{-1}$\,sr$^{-1}$, D=7.4\,pc), Tycho's SNR ($\Sigma$= 1.32$\times10^{-19}$\,W\,m$^{-2}$\,Hz$^{-1}$\,sr$^{-1}$, D=7.4\,pc) and Kepler's SNR ($\Sigma$= 3.18$\times10^{-19}$\,W\,m$^{-2}$\,Hz$^{-1}$\,sr$^{-1}$, D=8.18\,pc; \cite{2013ApJS..204....4P}).

\section{Conclusions}
 \label{sec:con}
In this paper, we produced new radio continuum images for the young LMC SNR \n\ using ATCA CABB data and compared them with pre CABB images. Our new images are of higher resolution and more sensitive than previous images obtained by \citet{dm}. 

We estimated \n\ radio spectral index of --0.75$\pm$0.01 and found its shape to be concave-up. We suggest that the most likely reason is the NLDSA effects or presence of two different population of ultra-relativistic electrons.

We found that the radio morphology for this SNR is asymmetric which is different from other young type\,Ia LMC SNRs such as J0509--6731 or LHG\,26, which exhibit a circular morphology.

We place an upper limit on the expansion velocity of the SNR shell and found our value to be consistent with that determined for higher spatial resolution using X-ray data. We detect that the SNR \n\ shows localized linear polarisation, especially in the north-west side of the remnant. The average fractional polarisation for \n\ at 5500\,MHz is 8$\pm$1\%, which is slightly higher than the value found by \cite{dm} due to higher sensitivity. We also estimate the $P$ value at 9000\,MHz to be 13$\pm$3\%. Interestingly, we found unpolarised clumps towards the south-west bright region (Figure~\ref{Fig4}) which might be a relic of the initial explosion. Our new $\alpha$ value for \n\ is consistent with that of previous studies of --0.75$\pm$0.01 and inline with other young type\,Ia SNRs.   

In order to estimate the strength of the magnetic field for this SNR, we calculated RM at three frequencies (4817, 5500, and 6183\,MHz). The mean value of RM was 200$\pm$20\,rad\,m$^{-2}$ and we estimated the magnetic field strength for this SNR to be $\sim$16.4\,$\mu$G. We also estimate the equipartition field to be $\sim$235\,$\mu$G with a minimum energy of E$_{\rm min}$=6.3$\times10^{48}$\,erg.

When comparing \n\ with other similar type\,Ia SNRs we suggest that the SD types could be somewhat asymmetrical in appearance while DD are well and circularly shaped.

\acknowledgments
The Australian Compact Array is part of the Australian Telescope which is funded by the Commonwealth of Australia for operation as National Facility managed by CSIRO. This paper includes archived data obtained through the Australia Telescope Online Archive (http://atoa.atnf.csiro.au). We used the \textsc{karma} and \textsc{miriad} software packages developed by the ATNF.

\bibliographystyle{spr-mp-nameyear-cnd}
\bibliography{biblio-u1}

\end{document}